\documentclass[prd, onecolumn,aps,preprintnumbers,letterpaper,floatfix,showpacs]{revtex4}
\usepackage[total={6in, 8in}]{geometry}
\usepackage{bm, graphicx, color}
\usepackage{amsmath,amssymb,amsfonts}
\usepackage{wasysym, hyperref,subfigure,wrapfig,upgreek}
\newcommand{\ddd}{\displaystyle}

\newcommand{\LL}[1]{\Lambda_{#1}}
\newcommand{\KK}{\kappa}
\newcommand{\sinc}{\mathrm{sinc}}
\newcommand{\DD}{D}
\newcommand{\EE}{W}
\newcommand{\vEE}{\vec W}
\newcommand{\hide}[1]{ }
\newcommand{\Pn}[1]{S_{n_{#1}}}
\newcommand{\vev}[1]{\langle #1 \rangle}

\begin{document}

\title{Searching for stochastic background of ultra-light fields with atomic sensors}

\author{Tigran Kalaydzhyan}\email{tigran.kalaydzhyan@jpl.nasa.gov}
\affiliation{Jet Propulsion Laboratory, California Institute of Technology,\\ 4800 Oak Grove Dr, MS 298, Pasadena, CA 91109, U.S.A.}
\author{Nan Yu}
\affiliation{Jet Propulsion Laboratory, California Institute of Technology,\\ 4800 Oak Grove Dr, MS 298, Pasadena, CA 91109, U.S.A.}

\date{\today}

\begin{abstract}
We propose a cross-correlation method for the searches of ultra-light fields, in particular, with a space network of atomic sensors. The main motivation of the approach is cancellation of uncorrelated noises in the observation data and unique pattern the fields leave on the cross-spectrum, depending on their nature (i.e., scalar, vector or tensor). In particular, we analytically derive a dependence of the cross-spectrum on the angle between two pairs of detectors. We then confirm obtained angular curves with a numerical simulation. We apply the method to the detection of dark matter and gravitational waves.

\end{abstract}

\pacs{95.35.+d, 06.20.fb, 04.80.Nn}

\maketitle

\section{Introduction}

Search for new fields has been the central part of modern high-energy physics. While most of the effort is concentrated in the mass range of GeVs and above~\cite{Patrignani:2016xqp}, where the new physics can be tested with particle accelerators and cosmic rays, relatively little is done in the sub-eV region. In that region, most active are the axion studies~\cite{Graham:2015ouw}, usually spanning the interval of $\upmu$eV-meV. Fields with even lower masses have gained more attention only very recently, mostly, as possible candidates for the dark matter (DM), see Refs.~\cite{Graham:2015ifn, Kalaydzhyan:2017jtv} for an overview. Large Compton wavelengths for such fields ($\lambda > 1\, \mathrm{m}$ for $m < 1\upmu$eV) require a methodology beyond the traditional particle physics but can be tested with atomic physics experiments and/or large scale experiments in space.
One example is a comparison of frequencies of two atomic clocks being affected by slowly oscillating scalar field background. Another would be a test of the weak equivalence principle with two isotopes in free fall, as an additional acceleration might be induced by a new ultra-light vector field. Example of a large-scale experiment would be a fifth-force type of searches, seeking for anomalies in the trajectories of celestial bodies and spacecrafts~\cite{Fukuda:2018omk} or searches for timing anomalies in the GPS data~\cite{Roberts:2017hla}. For an overview of possible spatial configurations of ultra-light fields, see Ref.~\cite{Kalaydzhyan:2017jtv} and Refs. therein. We recall that the current literature provides the following most common configurations of the fields: waves with the frequency at the field mass~\cite{VanTilburg:2015oza, Arvanitaki:2014faa, Hees:2016gop}, clumps (e.g., topological defects~\cite{Derevianko:2013oaa, Stadnik:2014cea, Wcislo:2016qng}), caustics~\cite{Prezeau:2015lxa} and simple static distributions~\cite{Leefer:2016xfu}. 

In this paper, we consider a new case: a stochastic background of waves of light fields, which we propose to measure by means of a network of precise atomic sensors. Some of the past methods assumed that the entire energy density of DM is carried by a monochromatic (or quasi-monochromatic~\cite{Derevianko:2016vpm, Budker:2013hfa, Arvanitaki:2017nhi, Foster:2017hbq}) wave with a frequency fixed at $m_\phi/2\pi$. However, if the total energy density is distributed over a range of frequencies, then the limits summarized in Ref.~\cite{Kalaydzhyan:2017jtv} will be significantly reduced. Therefore, it makes sense to put limits not only on the DM couplings, but also on the spectrum of DM excitations. Stochastic background of waves may appear in the context of Bose-Einstein Condensate (BEC) and superfluid dark matter models~\cite{Mocz:2017wlg, Berezhiani:2015pia, Berezhiani:2015bqa}. Technically, it may also be easier to search for a stochastic background of waves rather that single waves, because one does not have to form a bank of templates and the signal-to-noise ratio can be substantially improved by increasing the observation time (we discuss it later in the paper). We first analyze the cases of scalar and vector fields in the context of DM, and then also show how similar methodology can be used for tensor fields in the context of gravitational waves.

\section{Stochastic scalar fields}
In this section, we consider a new hypothetical scalar field, which we, for convenience, associate with DM to be able to refer to the existing literature.
The scalar field $\phi$ with mass $m_\phi$ will be coupled linearly to the Standard Model (SM) operators (see, e.g., dilaton dark matter studies~\cite{Damour:2010rp, Arvanitaki:2014faa, Hees:2016gop}),
\begin{align}
 &\mathcal{L}_{int} = \phi \left[ \frac{1}{4 e^2 \LL{\gamma}}  F_{\mu\nu}F^{\mu\nu} - \frac{\beta_{3}}{2g_{3} \LL{g}}G_{\mu\nu}G^{\mu\nu}-\sum_{f=e, u, d}\left(\frac{1}{\LL{f}}+\frac{\gamma_{m_f}}{\LL{g}}\right) m_f \bar \psi_f \psi_f\right],\label{Lagrangian}
\end{align}
where $F_{\mu\nu}$ and $G_{\mu\nu}$ standard electromagnetic and gluon field strength tensors, $\beta_{3}$ is the beta-function of the gauge coupling $g_{3}$, $\gamma_{m_f}$ is the anomalous dimension of the fermion (electron, u-, d-quark) mass operator and we use natural units, $\hbar = c = 1$. Parameters $\LL{a}$ have dimension of mass and play a role of (unknown) inverse coupling constants. For other possible couplings, see Refs.~\cite{Graham:2015ifn, Olive:2001vz, Olive:2007aj, Kalaydzhyan:2017jtv}. 
The chosen interaction Lagrangian, Eq.~(\ref{Lagrangian}), introduces local changes in values of fundamental constants, such as~\cite{Damour:2010rp}
\begin{align}
 &\frac{\delta \alpha}{\alpha} =  \frac{\phi}{\LL{\gamma}},\qquad
\frac{\delta m_f}{m_f} = \frac{\phi}{\LL{f}},\qquad \frac{\delta \Lambda_{QCD}}{\Lambda_{QCD}} =\frac{\phi}{\LL{g}},\label{variations}
\end{align}
where $\alpha$ is the fine-structure constant and $\Lambda_{QCD}$ is the QCD scale.
These variations can be studied by the changes in the atomic clock frequency\cite{Flambaum:2004tm}
\hide{
\begin{align}
\frac{\delta (\nu/\nu_0)}{\nu/\nu_0} = \frac{\delta V}{V},\qquad
V = \alpha^{K_\alpha}\left(\frac{m_q}{\Lambda_{QCD}}\right)^{K_{q}}\left(\frac{m_e}{\Lambda_{QCD}}\right)^{K_{e}},\label{response}
\end{align}
}
\begin{align}
\nu \propto\alpha^{K_\alpha}\left(\frac{m_q}{\Lambda_{QCD}}\right)^{K_{q}}\left(\frac{m_e}{\Lambda_{QCD}}\right)^{K_{e}},\label{response}
\end{align}
where 
$m_q = (m_u+m_d)/2$ and exponents $K_a$ are tabulated for the most common types of atomic clocks, see Refs.~\cite{Flambaum:2004tm, Flambaum:2008kr, Arvanitaki:2014faa} and references therein. As an example, for ${}^{133}$Cs, $K_{\alpha}=2.83$, $K_{q}=0.07$, $K_e = 1$. For optical clocks, only $K_{\alpha}$ is nonzero, i.e., the clocks are only sensitive to the changes in the fine structure constant. Using (\ref{variations}), we obtain
\begin{align}
\ddd\frac{\delta \nu}{\nu} = \phi \left[ \frac{K_\alpha}{\LL{\gamma}}+\frac{K_{q}}{\LL{q}}-\frac{K_{q}+K_{e}}{\LL{g}}\right], \quad \mathrm{where}~~ \LL{q} \equiv \frac{\LL{u} \LL{d} (m_u + m_d)}{m_u \LL{d} + m_d \LL{u}}.\label{freqshift}
\end{align}

For our study, it will be crucial that the fractional frequency variation depends linearly on the scalar field strength $\phi$, allowing us to express the Fourier transform of the frequency variation through the Fourier transform of $\phi$. This, together with the general requirements, such as the gauge- and Lorentz-invariance, is the main motivation behind the chosen form of the Lagrangian~(\ref{Lagrangian}). There exist strict experimental limits on the parameters $\LL{a}$ based on this (and similar) methods, see Refs.~\cite{Hees:2016gop, VanTilburg:2015oza, Kalaydzhyan:2017jtv}. However, these limits are based on the assumption of DM wave being monochromatic and can be relaxed in the case of a more general spectrum of DM waves. We refer the reader to the Appendix A for a relevant example. It is important to emphasize that the limits on $\Lambda_a$ should be always considered in the context of chosen DM configurations (e.g., waves, lumps, constant field, etc.). Since stochastic DM backgrounds were not considered before, we do not contrast possible $\Lambda_a$ sensitivities of our method to the projected sensitivities or ruled out regions of parameter space attributed to other methods.

\begin{figure}[t]
\centering
\subfigure[\label{Slide2}]{\includegraphics[width=6cm]{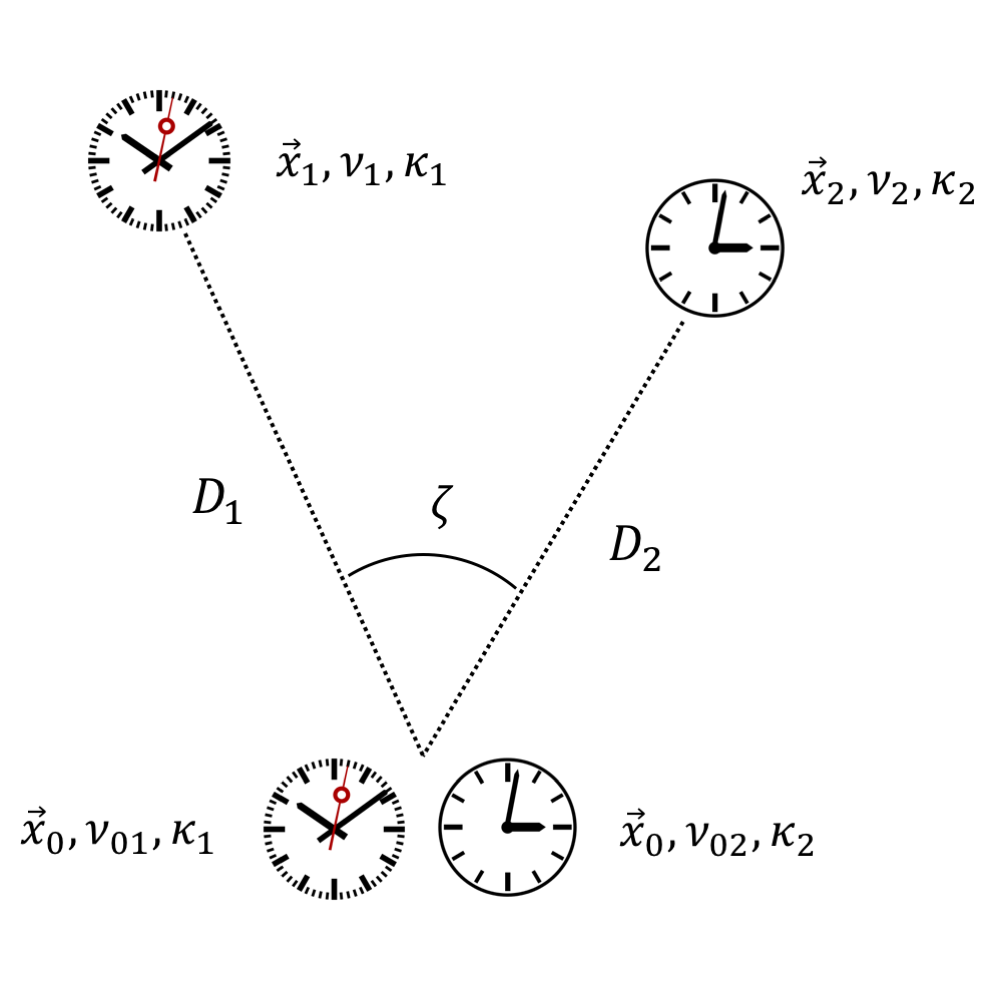}}\subfigure[\label{Slide3}]{\includegraphics[width=6cm]{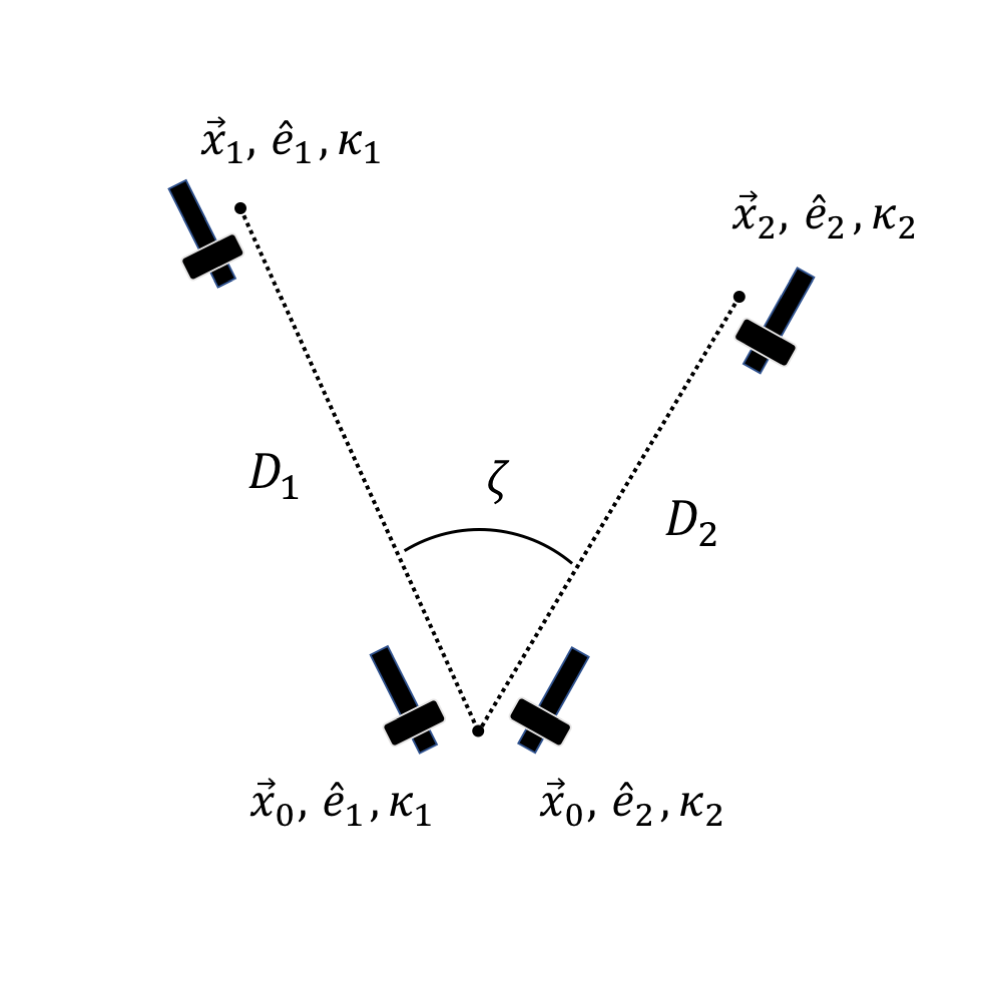}}
\caption{(a) Triangle-shaped configuration of two pairs of identical clocks.  Frequencies and sensitivities to DM are denoted by $\nu_i$ and $\KK_i$, respectively. (b) Triangle-shaped configuration of two pairs of identical atom interferometers pointing in radial direction (each direction is denoted by $\hat{e}_i$). Length of the first and the second arm is denoted by $D_1$ and $D_2$, respectively.}
\end{figure}

We consider a triangle-shaped configuration of atomic clocks with frequencies $\nu_i$, $i = 01, 02, 1, 2$, placed at positions $\vec x_i$, see Fig.~\ref{Slide2}. Each clock responds to the scalar field by a shift in the frequency, $\delta \nu_i$ such as $\delta \nu_i / \nu_i = \KK_i \phi$, cf. Eq.~(\ref{freqshift}). 
We will measure the difference in the relative frequency shifts, $X_1(t) = \delta \nu_{01} / \nu_{01} - \delta \nu_1 / \nu_1$ and $X_2(t) = \delta \nu_{02} / \nu_{02} - \delta \nu_2 / \nu_2$. In a real experiment, one compares absolute clock frequencies in pairs, so we will assume that the clocks are identical in each pair, and $X_i(t)$ represents the relative frequency difference between clocks in a pair with the gravitational redshift difference taken into account.
We have chosen two pairs of atomic clocks instead of three identical clocks because the common reference clock would introduce a possible unwanted correlated noise that can be mistaken for the signal.
We will be interested in the cross-spectrum between $X_1(t)$ and $X_2(t)$ and derive a universal dependency of this cross-spectrum on the angle between the pairs of atomic clocks. The approach for the derivation is similar to the calculation of the Hellings-Downs curve~\cite{Hellings:1983fr} for pulsar-timing arrays used in the gravitational wave searches. The scalar field in consideration does not have to be DM and can be any neutral light scalar field obeying the energy density limit on the hidden matter content in the solar system (or, more generally, in the confining volume of the performed experiment), see, e.g., Ref~\cite{Pitjev:2013sfa}.

\begin{figure}[t]
\centering
\includegraphics[width=7cm]{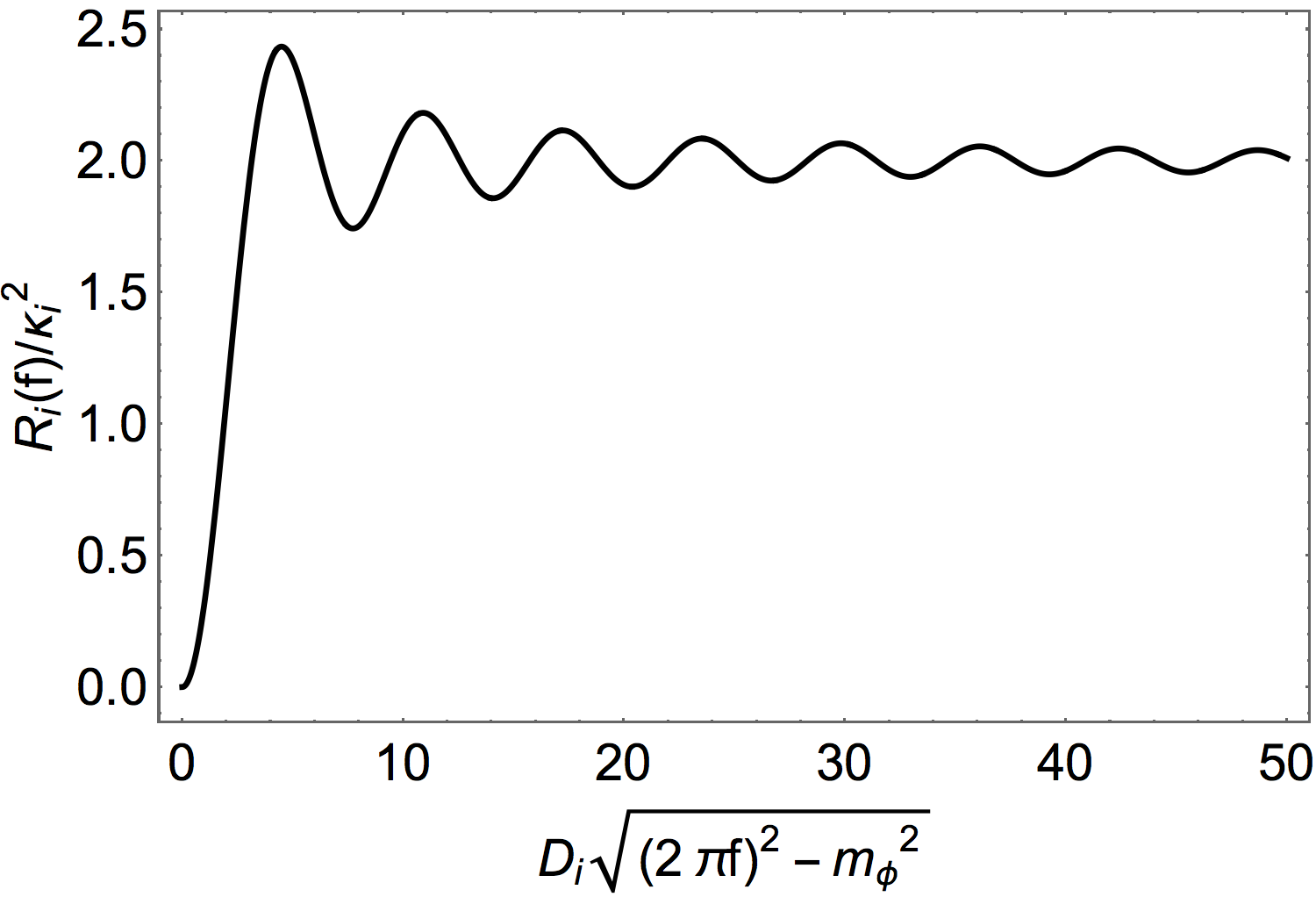}
\caption{Response function, $R_i(f)/\KK_i^2 = 2  \left[1-\sinc(k D_i) \right]$ for a pair of identical atomic clocks.\label{sinc}}
\end{figure}

In what follows, we assume that the data collection time $T$ is finite but large, giving the frequency resolution of the signal $\Delta f \sim 1/T$. The finite-time Fourier transform for all time-dependent quantities in this article is given by
\begin{align}
\tilde X(f) = T^{-1/2} \int_{-T/2}^{T/2} X(t) e^{-2\pi i f t} dt. 
\end{align}
We will be replacing the integration limits by infinity, whenever it does not lead to a confusion.
The power spectral density $S_X(f)$ of $X$ for one of the clock pairs is defined via $S_X(f) = |\tilde X (f)|^2$ and has the property (Parseval's theorem)
\begin{align}
\langle X^2(t) \rangle \equiv \lim_{T\to \infty} \frac{1}{T}\int_{-T/2}^{T/2} X^2 (t) dt = \int_{-\infty}^\infty S_X(f) df\,.\label{Parseval}
\end{align} 
We have chosen a two-sided version of the power spectral density to make it easier to change the order of integration, when needed. Notice that the given form of Fourier transform (together with the property of stationarity) leads to
\begin{align}
\langle \tilde X(f) \tilde X^* (f')\rangle = \frac{1}{T} \delta_T(f - f') S_X(f)\,,
\end{align}
where we defined the finite-time delta function as
\begin{align}
\delta_T(f)\equiv\int\limits_{-T/2}^{T/2} e^{-2\pi i f t} dt = T \sinc(\pi f t)\,.
\end{align}
If the two frequencies coincide, then $\delta(0)=T$ cancels the factor $T$ in the denominator and we recover the definition of the power spectrum.
Let us consider a stationary isotropic background of scalar field waves. We represent the scalar field as
\begin{align}
\phi(t, \vec x) = \int d^3 \vec k\, \tilde \phi (\vec k) e^{i (\vec k \cdot \vec x - 2 \pi f t)}\,,
\end{align} 
where the dispersion relation, in natural units, is $2\pi f = (m_\phi^2 + k^2)^{1/2}$ for massive matter waves or $f = \mathrm{const}\cdot k$ for gapless modes (such as phonons). Next, we introduce the cross-spectrum for two pairs of clocks,
\begin{align}
\displaystyle S_c(f, \zeta)=\int_{-\infty}^{+\infty}d\tau\, e^{-i 2\pi f \tau} \langle X_1(t) X_2 (t+\tau)\rangle = \langle \tilde X_1(f) \tilde X_2^* (f)\rangle\,.
\end{align}
We will be using two identities (the second one comes from the isotropic property of the background),
\begin{align}
\int\limits_{-\infty}^\infty df\, S_c(f, \zeta) = \langle X_1(t) X_2(t)\rangle, \qquad  \int\limits_{-\infty}^\infty df\, S_\phi(f) = 4\pi \int\limits_0^\infty dk \, k^2 \langle \tilde \phi(k) \tilde \phi^* (k)\rangle\,,
\end{align}
to calculate the cross-spectrum from the power spectrum of the scalar field,
\begin{align}
\int\limits_{-\infty}^\infty df\, S_c(f, \zeta) = \KK_1\KK_2 \int d^3 \vec k \langle\tilde \phi (\vec k)\tilde \phi^* (\vec k) \rangle \left(1 - e^{i \vec k \cdot (\vec x_1 - \vec x_0)} \right)\cdot\left(1 - e^{-i \vec k \cdot (\vec x_2 - \vec x_0)} \right) = \int\limits_{-\infty}^\infty df\, S_\phi(f) R_c(f, \zeta)\,,
\end{align}
where the response function $R_c(f, \zeta)$ is given by 
\begin{align}
R_c(f, \zeta) = \frac{\KK_1\KK_2}{4 \pi}\int_{S^2} d^2\Omega_{\hat k}\left(1 - e^{i \vec k \cdot (\vec x_1 -\vec  x_0)} - e^{-i \vec k \cdot (\vec x_2 - \vec x_0)} +e^{i \vec k \cdot (\vec x_1 - \vec x_2)} \right)\,.
\end{align}
By chosing the appropriate coordinate system (see, e.g., Ref.~\cite{Jenet:2014bea}), one can integrate each term in the sum and obtain
\begin{align}
R_c(f, \zeta) = \KK_1\KK_2 \left[ 1 - \sinc (k \DD_1) - \sinc (k \DD_2) + \sinc (k \DD_{12}) \right]\,,
\end{align} 
where $\DD_i \equiv |\vec x_i - \vec x_0|$,  $\DD_{12}\equiv  |\vec x_1 - \vec x_2|$, and $k = |\vec k|$. The response function $R_c(f, \zeta)$ relates the power spectrum of the scalar field fluctuations to the cross-spectrum of the measured signals,
\begin{align}
S_c(f, \zeta) = R_c(f, \zeta) S_\phi(f)\,.
\end{align}
If the scalar field $\phi$ is identified with DM, then its power spectrum density, $S_\phi(f)$, has the following integral characteristic,
\begin{align}
\rho_{\mathrm{DM}}\approx m_\phi^2 \langle\phi^2 \rangle = m_\phi^2 \int\limits_{-\infty}^{\infty}S_\phi(f) \, d f\,,
\end{align}
where $\rho_{\mathrm{DM}}$ is the average DM energy density in the vicinity of the experiment. A common model for the DM distribution in the Milky Way is a non-rotating isothermal spherical halo with velocities of the DM objects following Maxwell distribution. For a solar system experiment, it is believed that $\rho_{\mathrm{DM}}\approx 0.4\, \mathrm{GeV/cm^3}$~\cite{Olive:2016xmw} and that the DM objects are moving with (virial) velocities of $v_b \approx 270\, \mathrm{km/s}$ and the velocity dispersion is $\delta v_b \approx v_b$~\cite{Gelmini:2000dm}. The upper bound on $\rho_{\mathrm{DM}}$ is currently set to $\rho_{\mathrm{DM}} < 10^5\, \mathrm{GeV}/\mathrm{cm}^3$, based on positional observations of planets and spacecraft~\cite{Pitjev:2013sfa}.
Since the exact shape of $S_\phi(f)$ is not known {\it a priori}, from the practical point of view, it makes sense to eliminate it by normalizing the cross-spectrum with the power-spectrum of the signal from one pair of atomic clocks, $S_{X_i} = R_i(f) S_\phi(f)$, where $i=1, 2$, and $R_i(f) = 2 \KK^2 \left[1-\sinc(k D_i) \right]$, see Fig.~\ref{sinc}. It is clear from the plot that the sensitivity for a pair of identical atomic clocks reaches its maximum for the separation $\DD_i$ being larger than the scalar field wavelength. After the normalization of the cross-spectrum, the final result becomes
\begin{align}
F(f, \zeta)\equiv\frac{S_c(f, \zeta)}{S_{X_i}(f)} = \frac{\KK_1\KK_2}{\KK_i^2}\cdot\frac{1 - \sinc (k \DD_1) - \sinc (k \DD_2) + \sinc (k \DD_{12})}{2(1-\sinc(k \DD_i))}\,.\label{curve1scalar}
\end{align}
There are several natural limiting cases that can greatly simplify this expression due to the properties of the sinc function,
\begin{align}
F(f, \zeta) = \left\{ \begin{array}{l l} \ddd\frac{\KK_1\KK_2}{2\KK_i^2}, &  \DD_i \gg 1/k, \quad\DD_{12} \gg 1/k, \\&\\ \ddd\frac{\KK_1\KK_2}{\KK_i^2},&\DD_i \gg 1/k \gg \DD_{12},\\&\\ \ddd\frac{\KK_1\KK_2}{\KK_i^2}\frac{\DD_1 \DD_2}{\DD_i^2}\cos \zeta, & \DD_i \ll 1/k,\end{array} \right.\label{curve2scalar}
\end{align}
where $\zeta$ is the angle between $(\vec x_1 - \vec x_0)$ and $(\vec x_2 - \vec x_0)$. The second limit is geometrically restricted to a situation when $\DD_1 \approx \DD_2$ and $\zeta \approx \DD_{12}/\DD_i \ll 1$ or, in other words, when $2 \pi \DD_{12}$ is much smaller that the wavelength of the scalar field excitation.
It is important to notice that expressions in Eq.~(\ref{curve2scalar}) do not depend on the DM wave-vectors, frequencies and masses explicitly. Here we used the property $\sinc(a) \approx 0$ when $a \gg 1$ and $\sinc(a) \approx 1 - a^2/6$, when $a \ll 1$. Another nontrivial observation is that the first two limits give us nonzero constants~\cite{Jenet:2014bea}, while an uncorrelated noise would have given a vanishing cross-correlator. 

Eq.~(\ref{curve2scalar}) is the main result of this section. One can use it to identify the presence of a scalar signal in the cross-correlation data, either directly or by constructing an optimal filter to extract the signal from noise. In the described procedure, the frequency of the signal is bounded from below by several factors. First, the frequency should be larger than $m_\phi/(2\pi)$, due to the dispersion relation for the scalar field. Second, it should be larger than the inverse data collection time, $f \gg 1/T$, due to the property of the Fourier transform. Finally, it is limited by the time $T_\zeta \sim \Delta \zeta / \dot \zeta$ by which the angular distance between sensors changes by the value equal to the uncertainty in the angle $\zeta$, so $f \gg \dot \zeta /\Delta \zeta $. Value of $\Delta \zeta$ should be chosen such that there is enough recorded data for the extraction of the power spectrum density at the given frequency. 

The maximal frequency is limited by the Nyquist-Shannon-Kotelnikov theorem, $f_{max} = 1/(2 \Delta t)$, where $\Delta t$ is the time interval between consecutive measurements. Value of $\Delta t$ depends on the averaging interval giving desirable performance of the atomic sensors (typically, $\Delta t \geq 1$ s for high-performance atomic clocks). Frequencies larger than $f_{max}$ can be studied with the help of the temporal aliasing effect~\cite{Kalaydzhyan:2017jtv}, however, such analysis goes beyond the scope of this paper.

The method described here is general in nature and can be also applied beyond the use of atomic sensors. For instance, one could search for the long-wavelength curve, Eq.~(\ref{curve2scalar}), using instruments for the gravitational wave studies. First, one could analyze pulsar timing array data. When passing through a pulsar, the DM field can alter the neutron mass, size of the pulsar and hence its moment of inertia~\cite{Stadnik:2014cea}. This will lead to the variation of the pulsar rotation period. For the sensitivity estimates in the case of a single monochromatic DM wave, see Ref.~\cite{Graham:2015ifn}. Another method of DM detection with pulsar timing is presented in Ref.~\cite{Khmelnitsky:2013lxt}. Second, the DM wave could lead to the periodic displacement of test masses in LIGO and LISA experiments (for instance, LIGO mirrors), see the sensitivity estimates for $\LL{g}$ and a single monochromatic scalar wave in Ref.~\cite{Arvanitaki:2014faa}. Third, one could use binary pulsars for studies of stochastic backgrounds, see, e.g., Refs.~\cite{Blas:2016ddr, Hui:2012yp}.
 
\subsection*{Dark matter ``wind''}

DM direct detection experiments should take into account the possibility of DM moving with a constant velocity $\vec v_b$ with respect to the observer (DM ``wind"), see Ref.~\cite{Gelmini:2000dm} and refs. therein. Since the solar system moves through the galactic halo of the DM, such velocity may be given by the velocity of the Sun with respect to the galactic rest frame. If the solar system has its own DM halo, then the velocity of DM with respect to the near-Earth experiment would be given by the Earth orbital speed. 
In this section, we consider the effect of the constant motion of the configuration of detectors through the DM rest frame without making assumptions on the particular direction and absolute value of such constant velocity. Our goal is to see if there is a change in the angular part of cross-spectra when the velocities of DM waves are shifted by a constant vector $\vec v_b$. Another approach to the DM ``wind'' detection with atomic clocks is proposed in Ref.~\citep{Derevianko:2016vpm}.

For scalar excitations, due to the properties of the Fourier transform and Eq.~(\ref{Parseval}),
\begin{align}
\langle \tilde \phi (\vec k -\vec k_b)\tilde \phi^* (\vec k -\vec k_b)\rangle = S_\phi(f')\,,
\end{align}
where $\vec k_b \equiv m_\phi \vec v_b$, $S_\phi$ is the isotropic power spectrum and $f' = (m_\phi^2 + (\vec k - \vec k_b)^2)^{1/2}/(2\pi)$ is the Doppler-shifted frequency of the stochastic excitations. The cross-spectrum will be given then by $S_c(f, \zeta) = (df'/df) S_\phi(f') R_c(f', \zeta)$, where the response functions $R=R(f', \zeta)$ can be calculated in terms of the shifted wave-vectors, $\vec k =\vec k_b + \vec k'$,
\begin{align}
R_c(f', \zeta) = \frac{\KK_1 \KK_2}{4 \pi}\int_{S^2} d^2\Omega_{\hat k'}\left(1 - e^{i \vec k' \cdot \vec x_1} - e^{-i \vec k' \cdot \vec x_2} +e^{i \vec k' \cdot (\vec x_1 - \vec x_2)} \right)\,,
\end{align}
where we consider the configuration given by Fig.~\ref{Slide2}, and $\vec x_0 = 0$, for the sake of simplicity. After the integration, the expression becomes
\begin{align}
R_c(f', \zeta) = \KK_1 \KK_2 \left[ 1 - \sinc (k' \DD_1) - \sinc (k' \DD_2) + \sinc (k' \DD_{12}) \right]\,,
\end{align}
giving us familiar results (with the only difference of $k$ being replaced by $k'$),
\begin{align}
F(f, \zeta) = \left\{ \begin{array}{l l} \ddd\frac{\KK_1\KK_2}{2\KK_i^2}, &  \DD_i, \DD_{12} \gg 1/k',\\&\\ \ddd\frac{\KK_1\KK_2}{\KK_i^2},&\DD_i \gg 1/k' \gg \DD_{12},\\&\\ \ddd\frac{\KK_1\KK_2}{2\KK_i^2}\frac{\DD_1 \DD_2}{\DD_i^2}\cos \zeta, & \DD_i \ll 1/k',\end{array} \right.\label{angular_DMwind}
\end{align}
To reiterate the strategy, if the normalized cross-spectra give a constant (e.g., 1 or 1/2, for identical clocks) or cosine dependence on the angle between two pairs of atomic clocks, then the measured signal may be dominated by the presence of a new scalar field. In our derivation, we ignored the contribution of noise, assuming it to be weak. If the noise is strong, however, then one has to apply a different methodology, where the obtained analytic results, Eq.~(\ref{curve2scalar}) or Eq.~(\ref{angular_DMwind}), are used in the combination with a statistical analysis for the extraction of a weak signal from noise. We describe this approach in the next section.

\subsection*{Signal-to-noise ratio and statistical inference}

In this section, we provide the signal-to-noise ratio (SNR) for stochastic background measurements, an optimal filter, as well as the probability of the presence of the signal in the noise. This will allow us to estimate measurement parameters (such as the measurement duration) to achieve a given measurement sensitivity. Our derivations follow the gravitational wave detection methods~\cite{Allen:1997ad, Allen:1996vm, Romano:2016dpx}; for the basics of the signal extraction from noise see, e.g., Ref.~\cite{Zubakov}. We focus on the case of scalar fields, the generalization on other cases is straightforward. We consider the measured observable as a sum of a (weak) signal $\phi(t)$ and a noise $n(t)$, both having zero expectation values. For each pair of atomic clocks the noise is characterized by the (one-sided) power spectrum $\Pn{i}$, and the noises are assumed to be uncorrelated and stationary. In order to solve the problem of optimal filtering, we consider the signal to be
\begin{align}
S \equiv \frac{1}{T}\int\limits^{T/2}_{-T/2}dt \int\limits^{T/2}_{-T/2}dt'\, s_1(t) s_2(t') Q(t-t')\,. 
\end{align}
If the filter $Q(t-t') = \delta(t-t')$, then the signal is simply $S = \vev{\phi_1(t) \phi_2(t)}$. This signal can be, however, buried under noise and we need to construct a filter $Q(t)$ that will maximize SNR, eventually making it large enough at the expense of long observation time $T$. It will be shown that $\mathrm{SNR} \sim \sqrt{T}$ for our measurement. The Fourier space representation of the signal is
\begin{align}
S = \int\limits^{\infty}_{-\infty}df \int\limits^{\infty}_{-\infty}df'\, \delta_T(f-f') \tilde s_1^*(f)\tilde s_2(f') \tilde Q(f')\,,
\end{align}
where we assumed that $Q(\tau)$ decays quickly with $\tau \to \pm\infty$.
Next, statistical properties of the field amplitudes and the noise can be expressed as
\begin{align}
\vev{\tilde \phi_1^*(f) \tilde \phi_2(f')} = \frac{1}{T}\delta_T(f - f') F(f, \zeta) S_\phi (f)\,,\\
\vev{\tilde n_i^*(f) \tilde n_j(f')} = \frac{1}{2 T}\delta_T(f - f') \delta_{ij} \Pn{i} (|f|)\,. 
\end{align}
and, in addition to the property $\delta_T(0)= T$, lead to
\begin{align}
\vev{S} = \int\limits_{-\infty}^{\infty} df\, F(f, \zeta) S_\phi(f) \tilde Q(f)\,.\label{signal}
\end{align}
The noise is defined in the standard way, $N \equiv S - \vev{S}$. Assuming the noise contributions dominate the signal, we can write
\begin{align}
N \simeq \int\limits^{\infty}_{-\infty}df \int\limits^{\infty}_{-\infty}df'\, \delta_T(f-f') \tilde n_1^*(f)\tilde n_2(f') \tilde Q(f')\,,
\end{align}
and further obtain
\begin{align}
\vev{N^2} = \vev{S^2}-\vev{S}^2 \simeq \frac{1}{4 T} \int\limits_{-\infty}^{\infty}df\, \Pn{1}(|f|)\Pn{2}(|f|)|\tilde Q(f)|^2\,,
\end{align}
where we used a standard way of expressing the 4th moment through the covariances and took into account that $\int\limits_{-\infty}^\infty \delta_T^2 (f-f') df'=T$ at large $T$. Comparing this expression to Eq.~(\ref{signal}), one can see that $\mathrm{SNR}^2 = \vev{S}^2 / \vev{N^2}$ is maximized at
\begin{align}
\tilde Q(f) = \frac{F(f, \zeta)S_\phi(f)}{\Pn{1}(|f|)\Pn{2}(|f|)}
\end{align}
and gives
\begin{align}
\mathrm{SNR}^2 \simeq 8\, T \int\limits_0^\infty \frac{F^2(f, \zeta) S_\phi^2(f)}{\Pn{1}(f)\Pn{2}(f)} df\,.\label{SNR}
\end{align}
Notice that $\mathrm{SNR} \propto \sqrt{T}$ and the noise decreases with more data points collected, so, in ideal conditions, an arbitrarily small signal can be extracted from noise for long enough duration of the observation, which is, indeed, the power of the cross-correlation method.

In order to claim detection of a weak signal, one has to assume a certain shape of the spectrum $S_\phi(f)$, false alarm rate $\alpha$ and false dismissal rate $\beta$. Consider a set of $n$ statistically independent measurements $\left\{S_i\right\}$, each of duration $T$.
In order to test the null hypothesis (no signal of scalar waves in the data), one can form a random variable
\begin{align}
\hat X = \sqrt{n}\,\hat{\mathrm{SNR}} = \frac{\vev{S}}{\sqrt{\vev{(S-\vev{S})^2}}/\sqrt{n}}\,,
\end{align}
which is normally distributed with unit variance, in assumption of a large enough $n$. 
The Neyman-Pearson decision criterion (maximizing probability of the detection with fixed alarm rate $\alpha$) allows us to use it as a test statistic and choose the null hypothesis if $\hat X < X_{*}$, where $X_{*}$ is fixed by the choice of the false alarm rate,
\begin{align}
X_{*} = \sqrt{2}\,\mathrm{erfc}^{-1}(2 \alpha), \quad\mathrm{erfc}(x) \equiv \frac{2}{\sqrt{\pi}}\int\limits_{x}^\infty d\xi\, e^{-\xi^2}\,,
\end{align}
and reject the null hypothesis otherwise.
Here $\mathrm{erfc}^{-1}(x)$ is the inverse of the complementary error function. In other words, one can claim the presence of the DM signal of unknown amplitude if $\sqrt{n}\,\hat{\mathrm{SNR}} \leq \sqrt{2}\,\mathrm{erfc}^{-1}(2 \alpha)$.
At the next step, after the null hypothesis is rejected, i.e., assuming the signal is present in the data, the theoretical SNR required for the detection of the DM background in at least $(1-\beta)\times 100\%$ of measurements is given by~\cite{Allen:1997ad}
\begin{align}
\sqrt{n} \mathrm{SNR} \geq \sqrt{2}\left[\mathrm{erfc}^{-1}(2 \alpha)- \mathrm{erfc}^{-1}(2-2 \beta)\right]\,.
\end{align}
If the spectrum is flat, $S_\phi(f) = \bar S_\phi=\mathrm{const}$, in order to detect the signal, one would require the following minimal value of $\bar S_\phi$,
\begin{align}
\bar S_\phi \geq \frac{1}{2\sqrt{nT}}\left[\int\limits_0^\infty df\frac{F^2(f, \zeta)}{\Pn{1}(f)\Pn{2}(f)} \right]^{-1/2}\times\left[\mathrm{erfc}^{-1}(2 \alpha)- \mathrm{erfc}^{-1}(2-2 \beta)\right]\,.\label{minSpec}
\end{align}
The same technique can be applied to a network of more that two pairs of clocks, see Ref.~\cite{Allen:1997ad} discussing it in the context of gravitational wave detectors. For the anisotropic backgrounds and real data complications see, e.g., Ref.~\cite{Romano:2016dpx}.
For illustration purposes, let us consider a system of identical clocks in the short-wavelength regime ($F(\zeta) = 1/2$), with the scalar field background being localized around frequency $f_0$ in a narrow band $\Delta f$ and, for the particular statistics $\alpha=\beta=0.05$. Then the mentioned above expressions are simplified to 
\begin{align}
\mathrm{SNR} = \frac{\bar S_\phi \sqrt{2 T \Delta f}}{S_n(f_0)},\qquad \bar S_\phi^{5\%, 5\%} \geq \frac{2.33\, S_n(f_0)}{\sqrt{n T \Delta f}}\,,\label{flat_s}
\end{align}
where the power spectrum density is for the 5\% false alarm and false dismissal rates. Again, one see the advantage of the large number of measurement sessions and their long duration, with fixed levels of noise.

\begin{figure}[t]
\begin{center}
\includegraphics[width=6cm]{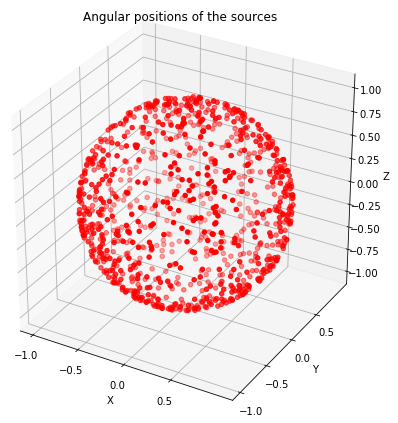}
\end{center}
\caption{\label{Sources} Angular positions of the sources}
\end{figure}

\subsection*{Numerical simulations}

\begin{figure}[t]
\centering
\subfigure[\label{signal_reference}]{\includegraphics[width=7cm]{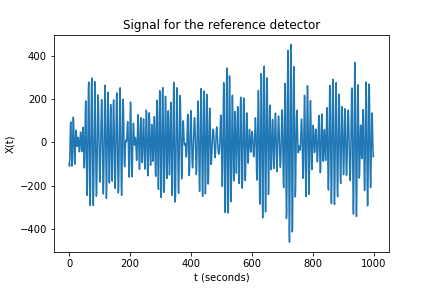}}
\subfigure[\label{spectrum_reference}]{\includegraphics[width=7cm]{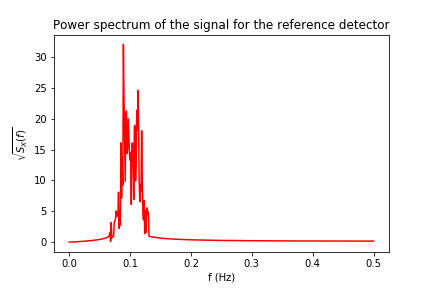}}
\subfigure[\label{cross_correlator}]{\includegraphics[width=7cm]{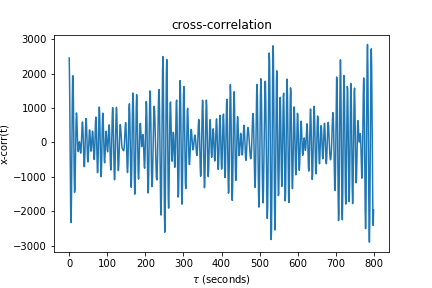}}
\subfigure[\label{spectrum_cross}]{\includegraphics[width=7cm]{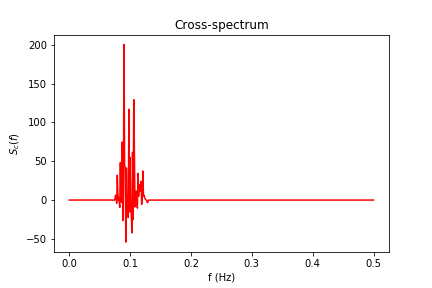}}
\caption{(a) Signal at the reference detector (clock); (b) Its spectrum; (c) Cross-correlator for two signals (two pairs of atomic clocks); (d) Cross-spectrum.}
\end{figure}

\begin{figure}[t]
\centering
\subfigure[\label{angular_long}]{\includegraphics[width=7cm]{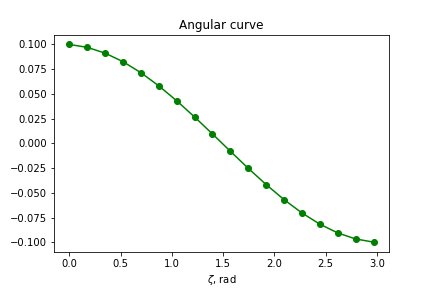}}
\subfigure[\label{angular_short}]{\includegraphics[width=7cm]{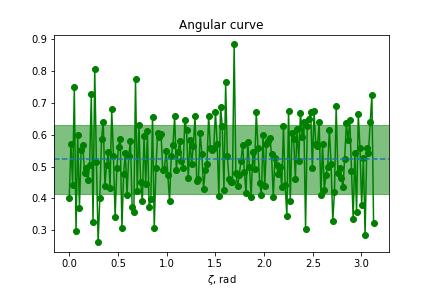}}
\caption{Angular curves for the (a) long-wavelength background; (b) short-wavelength background. The band shows $1\sigma$ confidence interval for the constant value of $F(\zeta)$.}
\end{figure}

In this section, we check validity of some of our results numerically. Necessity of the numerical test is due to the finite number of sources of DM waves, finite measurement time and number of measurement sessions (i.e., statistics). The numerical analysis provides some guidance with respect to the choice of measurement parameters and the impact of imperfections in the stochastic background. We choose $N=1000$ sources of DM waves randomly positioned in the sky (Fig.~\ref{Sources}) with random amplitudes, phases and random frequencies normally distributed around a given mean value.
For the angular positions, $P(\varphi) = (2 \pi)^{-1}$ and $P(\cos(\theta)) = 1/2$. Random amplitudes and phases mimic random distances to sources. We record the time-dependent signal with two pairs of detectors, assuming, at this stage, that there is no instrumental or any other kind of technical noise. The sampling time is chosen to be $\Delta t = 1$ s and the session length $T = 1000$ s. Signal recorded by one of the detectors is shown in Fig.~\ref{signal_reference} and its spectrum is presented in Fig.~\ref{spectrum_reference}. 
The cross-correlator for the two sets of data and the cross-spectrum are shown in Figs.~\ref{cross_correlator} and \ref{spectrum_cross}, respectively.
The angular curves are obtained within the frequency band given by the spread of the spectrum of the sources and shown in Figs.~\ref{angular_long} and \ref{angular_short}. The long-wavelength limit gives the identical result to the analytic formula, Eq.~(\ref{curve2scalar}), here $\DD_2 = 0.1\,\DD_1$ by choice. The short wavelength limit agrees with Eq.~(\ref{curve2scalar}) within at least one standard deviation.

\begin{figure}[t]
\centering
\subfigure[\label{noise_100}]{\includegraphics[width=7cm]{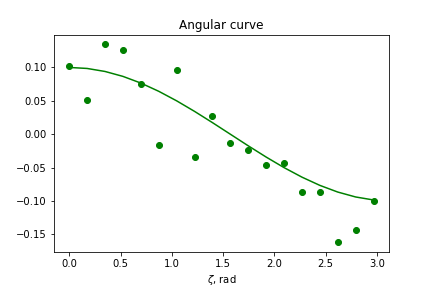}}
\subfigure[\label{noise_200}]{\includegraphics[width=7cm]{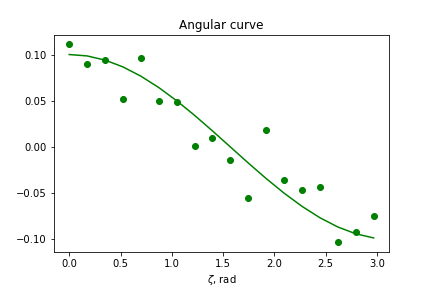}}
\subfigure[\label{noise_1000}]{\includegraphics[width=7cm]{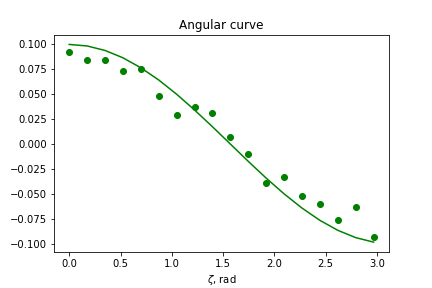}}
\subfigure[\label{noise_10000}]{\includegraphics[width=7cm]{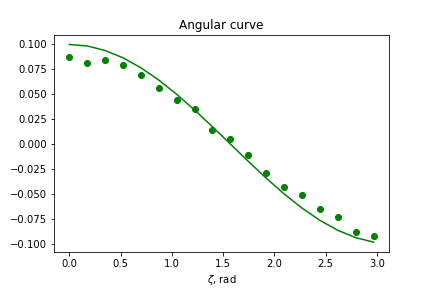}}
\caption{Angular curves for long-wavelength regime and weak noise. (a) T=100 s; (b) T=200 s; (c) T=1000 s; (d) T=10000 s. The solid line corresponds to the analytic result without detector noise. \label{noise_curves}}
\end{figure}

We also made preliminary measurements of the SNR with the optimal filter derived in the previous section. In the presence of a weak and strong white Gaussian noise, these measurements are consistent with the $\mathrm{SNR}\propto\sqrt{T}$ dependence. More precise quantitative statement requires additional computing power and will be made elsewhere. The effect of noise suppression with increase of observation time is demonstrated in Fig.~\ref{noise_curves}.


\section{Stochastic vector fields}
Another popular DM candidate is the massive $B-L$ vector field that exerts an additional force on neutral atomic species with different number of neutrons (e.g., different isotopes).
This addtional force can be probed with atom interferometer (AI) measurements in acccelerometer configurations. Such $B-L$ field detection has been mostly discussed in the context of experimental tests of the weak equivalence principle (WEP) and manifests itself in an additional relative acceleration between two particle species $a$ and $b$~\cite{Graham:2015ifn, Williams:2015ima},
\begin{align}
\Delta \vec a_{B-L}(t, \vec x) = \frac{g_{B-L}}{m_N} \left(\frac{Z_a}{A_a} - \frac{Z_b}{A_b} \right) \vEE (t, \vec x)\equiv \KK \vEE(t, \vec x)\,,
\end{align}
where $g_{B-L}$ is the coupling constant to the $B-L$ field $\vEE$, $Z_{a, b}$ and $A_{a, b}$ are the atomic numbers and weights, respectively, $m_N$ is the neutron mass, and we also introduced the constant $\KK$ for the sake of compactness. It has been proposed to measure such acceleration difference with the dual-species atom interferometers (AI)~\cite{Graham:2015ifn, Williams:2015ima}. For the review on principles of atom interferometry and use of AI as accelerometers, see, e.g, Ref.~\cite{Kasevich}.
Let us consider four AI in a triangle-shaped configuration, such that two of them are at the position $\vec x_0 =0$ and the other two are at the positions $\vec x_1$ and $\vec x_2$, respectively, see Fig.~\ref{Slide3}. All AI are oriented in the radial direction with respect to the center at $\vec x_0$. The quantities we are interested in are defined by the radial accelerations (along the lines connecting AIs in a pair),
\begin{align}
X_i(t) = \Delta a^r_{B-L}(t, \vec x_0) - \Delta a^r_{B-L}(t, \vec x_i)= \KK_i \hat{e}_i \cdot (\vEE (t, \vec x_0)-\vEE (t, \vec x_i))\,,
\end{align} 
where $\hat{e}_i$ are unit vectors directed along the AIs. We represent the vector field $\EE$ as
\begin{align}
\vec \EE(t, \vec x) = \int d^3 \vec k \sum\limits_{i=1,2,3} \tilde \EE_i (\vec k) \hat\epsilon_i(\hat k)\, e^{i(\vec k \cdot \vec x - 2\pi f t)}\,,
\end{align}
where $\hat\epsilon_i(\hat k)$ are polarization vectors, such as $\hat\epsilon_i(\hat k)\cdot \hat\epsilon_j(\hat k) = \delta_{ij}$ and the dispersion relation is $2 \pi f = (m_\EE^2 + k^2)^{1/2}$ with $m_\EE$ being the mass of the $B-L$ field~\footnote{Generalization to the massless case can be easily obtained by considering only two (transverse) polarizations. The final result for the angular curves will differ only by a numerical factor.}. We further consider a stochastic, isotropic and unpolarized background of vector waves characterized by the power spectrum density $S_{\EE}(f)$,
\begin{align}
\langle \tilde \EE_i (\vec k) \tilde \EE_j^* (\vec k) \rangle = \frac{1}{3}S_{\EE}(f) \delta_{ij}, \qquad \langle\vec\EE^2(t, \vec x)\rangle = \int\limits_{-\infty}^\infty S_{\EE}(f)\, df\,.
\end{align}
Notice the factor 3 in the denominator, which shows the number of physical polarizations. 
In analogy with the previous section, we calculate the cross-spectrum, $S_c(f, \zeta)=R_c(f, \zeta) S_\EE (f)$, where
\begin{align}
R_c(f, \zeta) = \frac{\KK_1 \KK_2}{4 \pi}\int_{S^2} d^2\Omega_{\hat k}\,\frac{1}{3}\sum\limits_{i=1,2,3}\left[\hat e_1\cdot\hat\epsilon_i(\hat k)\right]\left[\hat e_2\cdot\hat \epsilon_i(\hat k)\right]\left(1 - e^{i \vec k \cdot \vec x_1} \right)\cdot\left(1 - e^{-i \vec k \cdot \vec x_2 } \right)\,.\label{vecresponse}
\end{align}
To compute this integral we first fix the orientations of AI in the Cartesian $(x, y, z)$-coordinates,
\begin{align}
\hat e_1 = \hat z,\qquad \hat e_2 = \sin \zeta \,\hat x + \cos \zeta \,\hat z\,,
\end{align}
and then fix the orthonormal basis of polarization vectors in polar coordinates $(k, \theta, \varphi)$, 
\begin{align}
\hat \epsilon_1(\hat k) =& \cos \theta \cos \varphi \,\hat x + \cos \theta \sin \varphi \,\hat y - \sin \theta \,\hat z = \hat \theta,\\
\hat \epsilon_2(\hat k) =& -\sin \varphi \,\hat x + \cos \varphi \,\hat y = \hat \varphi,\\
\hat \epsilon_3(\hat k) =& \sin \theta \cos \varphi \,\hat x + \sin \theta \sin \varphi \,\hat y + \cos \theta \,\hat z = \hat k.
\end{align}
In the short-wavelength approximation, $\DD_i, \DD_{12} \gg 1/k$, we can neglect the exponents in the right-hand side of (\ref{vecresponse}) and obtain $R_c(f, \zeta)=\frac{\KK_1 \KK_2}{3}\cos \zeta$. If $\DD_i \gg 1/k,  \DD_{12} \ll 1/k$, then $R_c(f, \zeta)=\frac{2 \KK_1 \KK_2}{3}\cos \zeta$, similar to the previous section. Finally, in the long-wavelength limit, $\DD_i \ll 1/k$, taking into account that $\hat x_i = \hat e_i$ and expanding the exponent in series, we get $R_c(f, \zeta)=\frac{ \KK_1 \KK_2}{9}k^2 \DD_1 \DD_2 \cos^2 \zeta$. Considering the power spectrum for each individual pair of AI, $S_{X_i}(f) = R_i(f) S_\EE (f)$, we obtain $R_i(f) = \frac{2}{3}\KK_i^2$ if $\DD_i\gg 1/k$ and $R_i(f) = \frac{\KK_i^2}{15} (k \DD_i)^2 $ if $\DD_i\ll 1/k$. Finally, introducing the angular curve $F(f, \zeta)\equiv S_c(f, \zeta)/S_{X_i}(f)$ we summarize the results of this section:
\begin{align}
F(f, \zeta) = \left\{ \begin{array}{l l} \ddd\frac{\KK_1 \KK_2}{2\KK_i^2} \cos \zeta, &  \DD_i \gg 1/k,\quad \DD_{12} \gg 1/k, \\&\\ \ddd\frac{\KK_1 \KK_2}{\KK_i^2} \cos \zeta,&\DD_i \gg 1/k \gg \DD_{12},\\&\\ \ddd\frac{5}{3}\frac{\DD_1 \DD_2}{\DD_i^2}\frac{\KK_1 \KK_2}{\KK_i^2}\cos^2 \zeta, & \DD_i \ll 1/k,\end{array} \right.\label{curve1vector}
\end{align}
so the strategy of detecting the DM footprint in the data is to search for a cosine or cosine squared modulation of the normalized cross-spectra. 

There are several remarks in order. First, if one expects a contribution of the DM ``wind'', then the result repeats Eq.~(\ref{curve1vector}) with $k$ being replaced by $k'$ in the limits. Second, if the vector field excitation is gapless, then the number of physical polarizations is reduced to two and the problem resembles the case of electromagnetic stochastic background~\cite{Jenet:2014bea}. Third, if all conventional sources of acceleration were known (if, e.g., the setup was in deep space), can be controlled or are spectrally uncorrelated, then one could consider, for simplicity, four spatially separated single-species AI in configuration of long-baseline gradiometer~\cite{Chiow:2015faa}. In this case, the provided derivation of the angular curve can be repeated after subtraction of the all conventional accelerations. Finally, the triangular measurement configuration for $B-L$ vector field discussed above resembles a two-arm gravitational wave detector such as LIGO and LISA. We speculate that either of them can be used for the $B-L$ vector DM detection because of the differential acceleration between test masses and highly correlated laser noises in each arm.

\section{Stochastic tensor fields}

\begin{figure}[t]
\centering
\subfigure[\label{Figresp}]{\includegraphics[width=7cm]{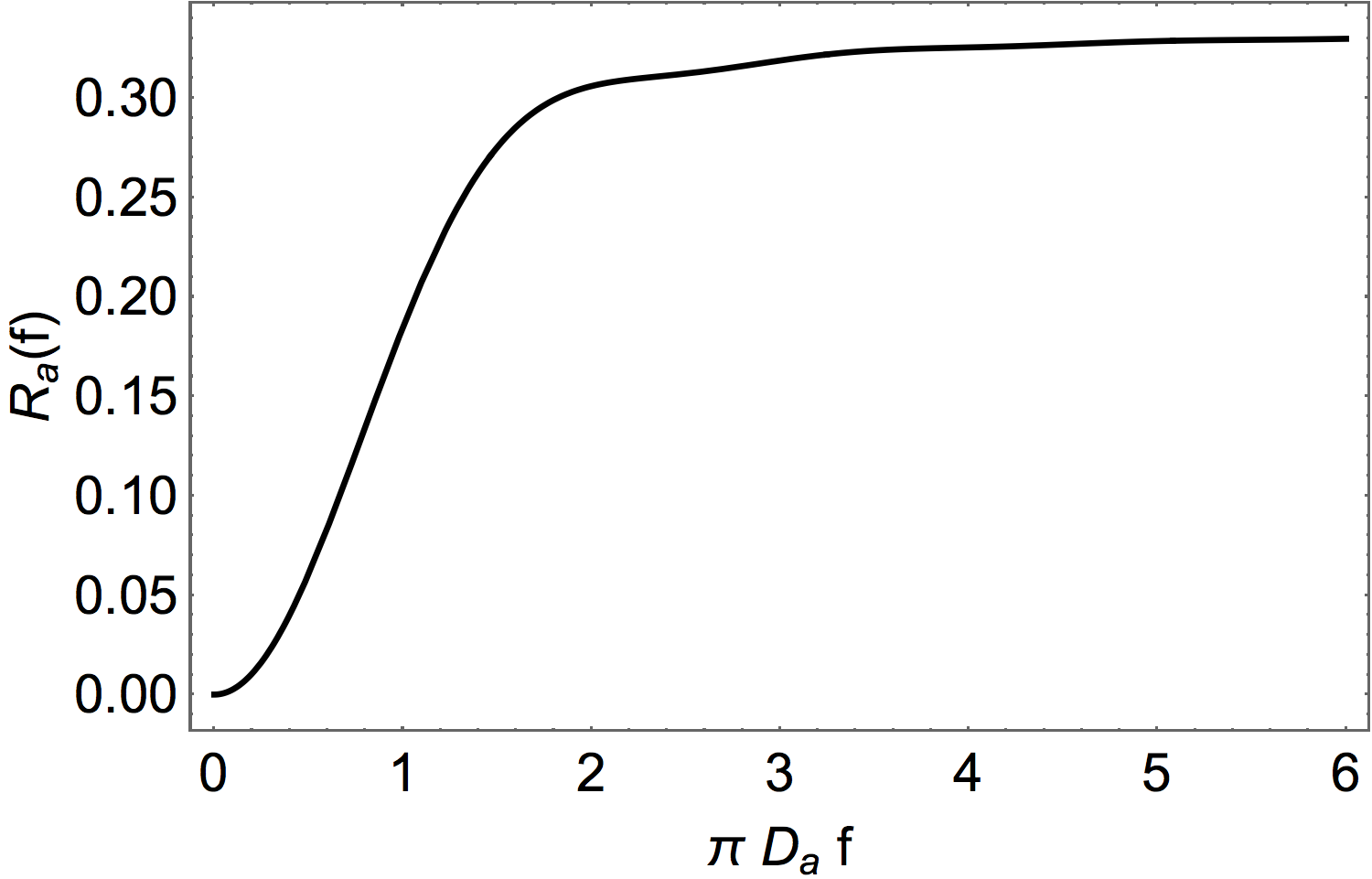}}
\subfigure[\label{curves}]{\includegraphics[width=7cm]{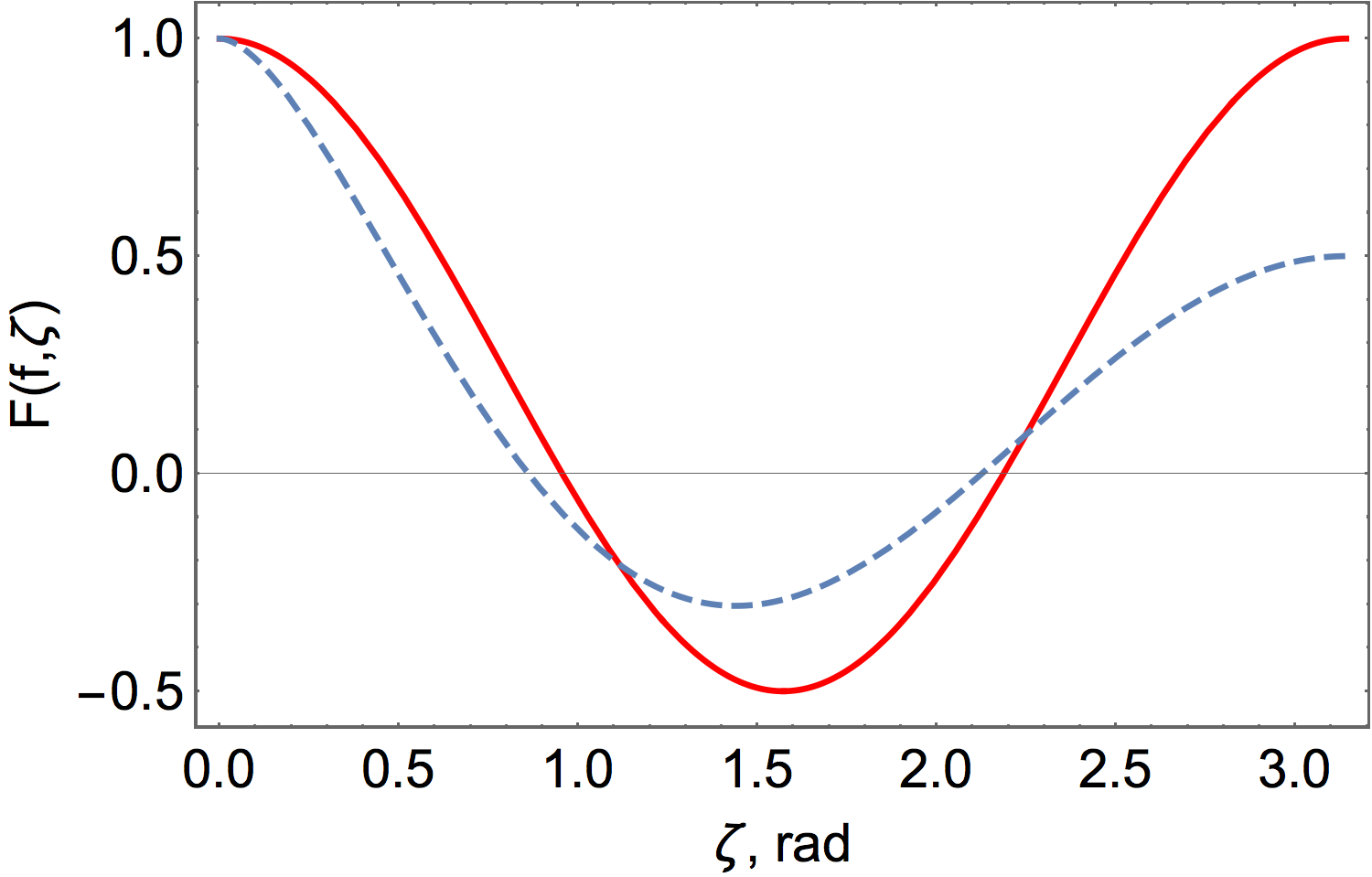}}
\caption{(a) The response function, Eq.~(\ref{singleresp}); (b) Normalized angular part of the cross-spectrum in the short-wavelength (dashed line) and long-wavelength (solid line) limits, Eq.~(\ref{curve1tensor})}
\end{figure}

In this section, we discuss an application of the method to the searches of isotropic unpolarized stationary gravitational wave (GW) background. Even though the method of cross-correlation of data from several detectors is widely used in the pulsar astronomy community (see, e.g., Refs~\cite{Allen:1997ad, Romano:2016dpx, Allen:1996vm}), we are not aware of the use of the method in direct detection experiments with atomic sensors. An example of the configuration is depicted in Fig.~\ref{Slide3}, where the atomic clocks are placed in spacecrafts or attached to celestial bodies in free fall. Frequency of an electromagnetic signal (e.g., laser) is locked to the frequency of one of the clocks in the pair and being compared with the frequency of the other clock, see, e.g., proposals in Refs.~\cite{Loeb:2015ffa, Kolkowitz:2016wyg, Vutha:2015aza}. Presence of GW will introduce a relative Doppler shift between the two frequencies. One can also use atom interferometers as gravitational wave detectors~\cite{Dimopoulos:2007cj, Yu:2010ss, Graham:2012sy, Graham:2016plp}, with test masses being atoms in excited and ground states. Even though in AI experiments the acceleration is measured rather than velocity, 
fundamentally, the observables are equivalent to the ones of the atomic clock detectors~\cite{Norcia:2017vwu}. 
We focus, for simplicity, on the case of clock comparison. The frequency difference due to the presence of the GW is denoted $X_a = \delta\nu_a/\nu_a$. Our goal, as before, is to extract the angular part of the cross-spectrum, $R_c(f, \zeta)$. The small perturbation of the metric around the Minkowski metric $\eta_{\mu\nu}$ is given by $g_{\mu\nu} =  \eta_{\mu\nu} + h_{\mu\nu}$. The metric perturbation $h_{\mu\nu}$ in a spatial transverse traceless gauge has $h_{0\mu}=0$ and can be represented by the plane wave expansion~\cite{Allen:1997ad}
\begin{align}
h_{ij}(t, \vec x) = \int_{-\infty}^{\infty}df \int_{S^2} d \hat \Omega\, e^{i 2\pi f (t - \hat \Omega\cdot\vec x)}\sum_{P=+, \times} h_P(f, \hat\Omega)e^P_{ij}(\hat \Omega)\,,
\end{align}
where $f$ is the frequency of the wave, unit vector $\hat \Omega$ points in the direction of the propagation of the plane-wave component. Dispersion relation for the gravitational waves is simply $2\pi f=k$. The polarization tensors can be defined through unit vectors $\hat n$ and $\hat m$ orthogonal to $\hat \Omega$,
\begin{align}
e^+_{ij}(\hat\Omega) = \hat m_i \hat m_j - \hat n_i \hat n_j, \quad e^\times_{ij}(\hat\Omega) = \hat m_i \hat n_j + \hat n_i \hat m_j\,,\label{polarizations1}
\end{align}
where in the standard spherical coordinates
\begin{align}
\hat \Omega &= (\sin \theta \cos \phi, \sin \theta \sin \phi, \cos \theta),\\
\hat m &= (\sin \phi, -\cos \phi, 0),\\
\hat n &= (\cos \theta \cos \phi, \cos \theta \sin \phi, -\sin \theta).\label{polarizations2}
\end{align}

We begin from considering an effect of a single plane wave propagating in direction $\hat\Omega$ on a signal sent between atomic clocks separated by distance $\DD_a$. The unit vector $\hat p_{(a)}$ is pointing from the observation point to the singnal source. In order to find the Doppler shift $X_a$, one can consider the null vector\cite{Hellings1981}
\begin{align}
\sigma^\mu_{(a)} = s^\mu_{(a)} - \frac{1}{2}\eta^{\mu\alpha}h_{\alpha\beta}s^{\beta}_{(a)}\,,
\end{align}
at the moments when the signal is emmited and received, with the unperturbed value given by $s^\mu_{(a)} = \nu (1, -\hat p_{(a)})$. The null geodesics can be found by solving the null condition $\sigma_{\mu (a)}\sigma^\mu_{(a)} = 0$ together with the standard condition $\sigma_\mu V^\mu_{(a)} = \mathrm{const}_{(a)}$, where $V^\mu_{(a)}$ are the Killing vectors for the perturbed geometry. The final result is given by
\begin{align}
X_{a}(t) = \frac{1}{2}\sum\limits_{P, i, j}\frac{\hat p^i_{(a)}  \hat p^j_{(a)} e^{P}_{ij}}{1+\hat \Omega \cdot \hat p_{(a)}}\left(h_P\left[t - \left(1+ \hat \Omega \cdot \hat p_{(a)}\right)\DD_a \right] - h_P[t] \right)\,,\label{dopplershift}
\end{align}
where we take into account the isotropic nature of the radiation and consider the perturbation amplitudes as functions of light-cone coordinates.
The power spectrum density for the induced Doppler shifts, $S_{X_a}(f) = \langle \tilde X_a(f) \tilde X_a^*(f) \rangle$
is given by averaging over time, directions $\hat \Omega$ and polarizations of the incoming radiation. It can be shown that $S_{X_a}(f) = R_a(f) S_h (f)$, where the GW power spectral density is defined by 
$\langle\tilde h_P(f)  \tilde h_{P'}^*(f) \rangle = \frac{1}{2}S_h(f) \delta_{P P'}$ 
and the response function is given by
\begin{align}
R_a(f) = \frac{1}{3}-\frac{1}{8 (\pi \DD_a f)^2} + \frac{\sin(4 \pi \DD_a f)}{32 (\pi \DD_a f)^3},\label{singleresp}
\end{align}
see Fig.~\ref{Figresp}. In a long wavelength limit, $2 \pi f \DD_a \ll 1$, the response function reduces to $R_a(f) = \frac{4 \pi^2}{15} (\DD_a f)^2$, while in the short wavelength limit it is simply $R_a(f) =1/3$.
By considering Eq.~(\ref{dopplershift}) in the Fourier space, one can derive the cross-spectrum, $S_c(f, \zeta) =\langle\tilde X_1(f)\tilde X^*_2(f) \rangle = R_c(f, \zeta) S_h(f)$, where
\begin{align}
R_c(f, \zeta) = &\frac{1}{4\pi}\int\limits_{S^2} d\hat \Omega \left(e^{2 \pi i f \DD_1 \left(1+\hat \Omega \cdot \hat p_{(1)}\right)} -1 \right)\left(e^{-2 \pi i f \DD_2 \left(1+\hat \Omega \cdot \hat p_{(2)}\right)} -1 \right) \nonumber\\
&\times\frac{1}{8}\sum\limits_{i, j,l, m, P} \frac{\hat p^i_{(1)}  \hat p^j_{(1)} e^{P}_{ij}}{1+\hat \Omega \cdot \hat p_{(1)}}\cdot \frac{\hat p^l_{(2)}  \hat p^m_{(2)} e^{P}_{lm}}{1+\hat \Omega \cdot \hat p_{(2)}}\label{longrc}
\end{align}
In order to perform the direction averaging for the radiation background, we chose the $z$-axis to be along $\hat p_{(1)}$ and $\hat p_{(2)}$ to have an angle $\zeta$ with respect to $\hat p_{(1)}$, similar to the previous section,
\begin{align}
\hat p_{(1)} = (0,0,1), \qquad \hat p_{(2)} = (\sin \zeta,0,\cos \zeta).
\end{align}
The angular dependency on $\zeta$ in the power spectrum density $S_c(f, \zeta)$ can be factorized in the short-wavelength limit, $\DD_a, \DD_{12} \gg 1/k$ and corresponds to the Hellings-Downs curve\cite{Hellings:1983fr, Anholm:2008wy} used in the pulsar timing studies, 
\begin{align}
R_c(f, \zeta) = \displaystyle \frac{1}{3}+ \frac{1}{2}(1-\cos\zeta)\left[ \ln\left( \frac{1-\cos\zeta}{2}\right) - \frac{1}{6}\right], \qquad \DD_a, \DD_{12} \gg 1/k\,.
\end{align}
To calculate this expression, one can neglect exponents in Eq.~(\ref{longrc}), since their arguments are quickly oscillating functions of spherical angles.  When $\DD_a \gg 1/k$, but $\DD_{12} \ll 1/k$, one will get the same result multiplied by factor 2, as in the previous sections.  In the opposite limit, $\DD_a \ll 1/k$, on can expand the exponents in series and obtain
\begin{align}
R_c(f, \zeta) = \frac{\pi}{8}\int\limits_{S^2} d\hat \Omega\, f^2 \DD_1 \DD_2 \sum\limits_{i, j,l, m, P} {\hat p^i_{(1)}  \hat p^j_{(1)}} {\hat p^l_{(2)}  \hat p^m_{(2)}  e^{P}_{ij} e^{P}_{lm}} = \frac{\pi^2 f^2}{15}\DD_1\DD_2 (1+ 3 \cos 2\zeta)\,,
\end{align}
which reproduces the response function of one pair of atomic clocks in this limit, when $\zeta=0$. One can notice that the angular function above is proportional to the second Legendre polynomial, $P_2(\cos \zeta)$, which one can expect from the expansion of the background by spherical harmonics~\cite{Bertotti1985, Gair:2014rwa}. Finally, introducing, again, the angular curve $F(f, \zeta)\equiv S_c(f, \zeta)/S_{X_ai}(f)$ we conclude this section,
\begin{align}
F(f, \zeta) = \left\{ \begin{array}{l l} \displaystyle 1+ \frac{3}{2}(1-\cos\zeta)\left[ \ln\left( \frac{1-\cos\zeta}{2}\right) - \frac{1}{6}\right], &  \DD_a \gg 1/k,\quad \DD_{12} \gg 1/k, \\&\\ \displaystyle 2 + (1-\cos\zeta)\left[ 3\ln\left( \frac{1-\cos\zeta}{2}\right) - \frac{1}{2}\right],&\DD_a \gg 1/k \gg \DD_{12},\\&\\ \ddd\frac{\DD_1 \DD_2}{4\DD_a^2}(1+3\cos 2 \zeta), & \DD_a \ll 1/k,\end{array} \right.\label{curve1tensor}
\end{align}
We plot the first expression and the third expression (with $\DD_1=\DD_2$) from that equation in Fig.~\ref{curves}. Sensitivity of a subsystem of a pair of atomic sensors is discussed in, e.g., Ref.~\cite{Kolkowitz:2016wyg}.

\section{Conclusions}

In this article, we proposed a general method of experimental measurements of ultra-light (or even massless) fields with atomic sensors. Such fields include various dark matter candidates (e.g., a dilaton or B-L field) and the gravitational field. We assumed that the fields form an isotropic stationary background and are being measured by two pairs of detectors in a triangular-shaped configuration (with or without variable angle). Depending on the nature of the fields (scalar, vector or tensor) they will leave a characteristic angular dependence on the cross-correlation of the data obtained from each pair. If the instrumental noise is weak, then the angular dependence of the cross-spectrum can be measured directly. If the noise is stronger than the expected signal, then the analysis should be done in a different way, that involves a statistical analysis with optimal filtering (we used the frequentist approach in our paper). Such analysis allows to recover the signal from a strong noise by performing long time observations. If the signal is not seen but is expected to be present, then one could put limits on the parameters of the model (such as DM couplings or characteristics of $S_\phi$). This, however, would require further assumptions on the shape of power spectrum density of the stochastic background (see, e.g., Eq.~(\ref{flat_s}) and Appendix A), which has to be motivated by the underlying theory of DM or GW generation. Comprehensive review of such theories is beyond the scope of this paper.
\vspace{6pt} 

\section*{Acknowledgements}
We are grateful to Slava Turyshev and Dmitry Duev for useful discussions. This work was performed at the Jet Propulsion Laboratory, California Institute of Technology, under a contract with the National Aeronautics and Space Administration. \textcopyright~2017 California Institute of Technology. Government sponsorship acknowledged.

\section*{Appendix A}
The advantage of the cross-spectrum analysis presented in this article is the suppression of uncorrelated noises from individual clocks through averaging and increase in sensitivity by moving the clocks apart. Certain limits, however, can be put by performing a much more simplified experiment. One could compare a single clock with another frequency standard (including another atomic clock), where the reference frequency standard has a much weaker or no dependence on the DM background. The clock instability in the time domain can be obtained from the power spectrum of the fractional variation of the clock frequency~\cite{Allan}. Assuming the DM is the dominating source of clock instability, the Allan variance for a clock with response coefficient $\KK$ will be given by
\begin{align}
\sigma^2_y(\tau) = 2\KK^2 \int\limits_0^{\infty} S_\phi(f) \frac{\sin^4(\pi \tau f)}{(\pi \tau f)^2} df\,.\label{Stosigma}
\end{align} 
The kernel in Eq.~(\ref{Stosigma}) is vanishing in the limit of very small and large $f$, so we can replace the integration limits by zero and infinity.
Extracting limits on the DM coupling directly from the clock instability data does not seem possible, due to the lack of knowledge on the DM power spectrum density and $\KK$. If, however, the DM spectrum corresponds to a white noise, $S_\phi(f)=\bar S_\phi=$ const, and $\sigma_y(\tau)\propto \tau^{-1/2}$, then for signal-to-noise ratio SNR=1, we have $\KK^2 \bar S_\phi = 2 \tau \sigma^2_y(\tau)$ -- a limit on the unknown combination of $\bar S_\phi$ and coupling $\KK$. As an example, for a ${}^{199}\mathrm{Hg}^+$ clock~\cite{NISTAlHg} this would mean a bound $\KK^2 \bar S_\phi\leq 3\times 10^{-29}\, \mathrm{Hz}^{-1}$ and, hence, $\bar S_\phi/\LL{\gamma}^2\leq 3\times 10^{-30}\,\mathrm{Hz}^{-1}$. From this expression, it is evident that the limit on $\LL{\gamma}$ can be lowered significantly (comparing to a monochromatic DM wave), if the power is smeared over a wide enough spectral band. Similar examples can be presented for other scales $\LL{a}$ by comparing, e.g., microwave clocks. Assuming that the clock instability is dominated by the DM effects leads to very conservative limits, that can be further improved by taking into account well-understood clock noise contributions.

\end{document}